\documentstyle[preprint,aps]{revtex}
\begin{document}
\input{psfig.sty}
\title{
Gluons as Goldstone Bosons when Flavor Symmetry is Broken Spontaneously}
\author{Nils A. T\"ornqvist}
\address{Department of  Physics,
POB 9, FIN--00014, University of Helsinki, Finland}
\date{5 March 1997}                 
\maketitle
\begin{abstract}
The mechanism where flavor symmetry of  the standard model
is broken spontaneously is discussed within a QCD model with
effective three-meson couplings. 
For sufficiently large coupling the model is  unstable with respect
to quantum loops from mesonic vacuum polarization.
It is argued that color and gluons 
naturally can account for the  Goldstone degrees
of freedom expected when flavor symmetry is spontaneously broken. 
\\
\vskip 0.05cm 
\noindent Pacs numbers:12.39.Ki, 11.30.Hv, 11.30.Qc, 12.15.Ff
\vskip 0.20cm 
\end{abstract}

As an introduction let me discuss a simple semiclassical
analogue to the mechanism which I am to discuss,
 which is easy to understand intuitively, without any formulas:
Consider classical balloons or balls, with fixed surface
area, inflated by hot air. When cooled the inside pressure 
can fall below the outside
pressure and the O(3) spherical balls must collapse spontaneously down
to at least O(2) ellipsoids\footnote{Deformed
nuclei which are common among the lantanides or actinides, e.g.$^{232}$Th,
are actually examples of this O(3)
symmetry breaking due to the balance of Coulomb and surface energies.}. 
Below the critical temperature the ellipsoids suddenly pick out  preferred
directions, the principal axes, and develop different moments of inertia.
The O(3) symmetry is broken in the shapes, but 
the O(3) symmetry  still remains in the sense that all rotated 
states of the collapsed system have the same energy. The directions of 
collapse (1,2,3 in Fig.1) are of course arbitrary.
This freedom of  rotating the ground 
state of the collapsed systems correspond to the Goldstone
degrees of freedom in a field theory with degenerate vacua.
But here "the vacuum"  (the surrounding air) need not break the symmetry,
only the shapes of the solutions break the spherical symmetry.

 
When   considered as a local symmetry ("collapsed
ellipsoids attached to each $x$-coordinate") the directions of collapse can 
be $x$-dependent.
 Broken flavor symmetry, which I shall discuss below,
 corresponds  to the broken O(3) symmetric shapes in this analogy and
the different masses within flavor multiplets to the
 unequal moments of inertia along the principal axes 
in Fig.1. The moments of inertia, 
are like the masses of a flavor multiplet. They are 
independent of $x$ as in  a broken global symmetry.
On the other hand  the  $x$-dependent freedom to rotate 
the ellipsoids, remains as an exact unbroken local symmetry. 
But both symmetries are defined within the same internal space. 
  
In QCD color and flavor have a kind of complementary role: 
At short distances inside hadrons, the color and gluonic degrees of
freedom are crucial, while the interactions are flavor independent. On
the  other hand  at large distances the color degrees of freedom 
are absent, since hadrons are color singlets, 
while flavor symmetry and its breaking is evident in  the mass spectrum.
Conventionally \cite{leut} 
one breaks flavor symmetry by adding, by hand, effective
non-degenerate quark masses to the QCD Lagrangian,  whereby             
the pseudoscalars obtain (small)  masses
and the degeneracy of all flavor  multiplets is split.
Most of the chiral quark masses are assumed to come from a
short distance regime, where  weak interactions, and the Higgs mechanism
are relevant.

In two recent preprints \cite{NAT3} I discussed a new mechanism 
where flavor symmetry
was broken spontaneously by  quantum effects, and where no quark mass
splitting term is needed in the Lagrangian. Trilinear 
meson couplings can lead to unstable 
self-consistency equations for
mesons dressed by the clouds of the same mesons. 
A natural question which a arises in this context of spontaneous flavor
symmetry breaking is: Where are the Goldstone bosons and the Goldstone degrees
of freedom in such models.
In this paper I  suggest a natural answer to this question 
within a scalar QCD model, when both color and flavor
obey the same SU3 symmetry group.

Let a nonet of real 
meson fields be described by a 3$\times$3 matrix ${\bf \Phi}(x)$ such
that an individual meson field is  $\Phi^\alpha ={\rm Tr} [{\bf \Phi
\Lambda}^{\alpha \dagger } ]$,
or inversely ${\bf \Phi}=\sum_\alpha \Phi^\alpha {\bf \Lambda}^\alpha$.
Here ${\bf \Lambda}^\alpha $ is a complete orthonormal set of 
nine real flavor matrices, normalized such that
${\rm Tr} [ {\bf \Lambda}^\alpha {\bf \Lambda}^{\beta \dagger}
]=\delta_{\alpha\beta }$.
The simplest  choice
for these, which we denote ${\bf \Lambda}^{ij}$,
where $i,j$ run from 1 to $N_f$ and for which all matrix elements are 0 
except the (i,j)'th matrix element which is 1 
($[{\bf \Lambda}^{ij}]_{m,n}=\delta_{im}\delta_{jn}$).
 The flavorless states
$u\bar u$, $d\bar d$,  $s\bar s$, which are represented by $\Lambda^{ii}$,
  can of course mix through an orthogonal matrix $\Omega$, such 
that the diagonal matrices are replaced by 
$\sum_j\Omega_{ij}{\bf \Lambda}^{jj}$. The mixing matrix $\Omega$ will
be determined by our self-consistency equations. 
(With isospin exact $\Omega$ mixes
 ${\bf \Lambda}^{11}$ and ${\bf \Lambda}^{22}$ to  $({\bf
\Lambda}^{11}\pm {\bf \Lambda}^{22})/\sqrt 2$).

We can write for scalar QCD, with two degenerate 
scalar nonets ${\bf  \Phi}_C$, $C=+,-$ of opposite charge conjucgation $C$
the flavor symmetric Lagrangean:
\begin{equation}
{\cal L} = -\frac 1 4 F^a_{\mu\nu}F^{a\mu\nu} + \sum_C {\rm Tr}[{\bf D}_\mu
{\bf \Phi}_C {\bf D}_\mu {\bf \Phi}_C^\dagger ] 
-m_0^2\sum_C{\rm Tr}[{\bf \Phi}_C{\bf \Phi}_C^\dagger ]+
{\cal L}_{int} \ ,
\end{equation}
where the first term is the usual pure gluonic term and
${\bf D}_\mu = \partial_\mu- ig{\bf T}^a A^a_\mu\ $ 
is the usual covariant derivative
and for ${\cal L}_{int}$ we chose

\begin{equation}
{\cal L}_{int}=g_F{\rm Tr}[{\bf \Phi_+^\dagger\Phi_+\Phi_-}]_- +
               g_D{\rm Tr}[{\bf \Phi_+^\dagger\Phi_-\Phi_-}]_+
                \ .
\end{equation}

We need the two nonets with opposite $C$ quantum numbers in order to have 
both F- and D- type couplings present, as in every more realistic model.
(E.g. if one includes the ground state mesons, one has $C=+$
pseudoscalars (P) and $C=-$ vector mesons (V) mesons, with PPV and PVV
couplings of both types.) Of course, 
we have to define the model as an effective
theory with a flavor independent cutoff $\Lambda$ 
in order to render it finite.
                                 
The color gauge group is unbroken, i.e., none of the scalar  fields develop a
vacuum expectation value, since the mass terms assumed $m^2_0 {\rm
Tr}[{\bf \Phi_\pm \Phi_\pm }^\dagger]$ have  a quadratic minimum at the
origin, and $<\Phi_{\pm}^\alpha  >=0$. 
Thus the gluons remain massless as they should. 
Under a gauge transformation 
${\bf U}(x)=exp(-i{\bf T}^a\theta^a(x))$ the fields
${\bf \Phi} $, the matrices ${\bf \Lambda}^\alpha$, and the covariant
derivative transform in a covariant way:
\begin{eqnarray}
{\bf \Phi}'_C &=&{\bf U}(x) {\bf \Phi}_C{\bf U}^{-1}(x) \ , \\
{\bf \Lambda}^{' \alpha }  &=&{\bf U}(x) {\bf \Lambda}^\alpha
{\bf U}^{-1}(x) \label{lam} \ , \\
{\bf D_\mu \Phi }'_C &=&{\bf U}(x) {\bf D_\mu \Phi }_C{\bf U}^{-1}(x) \ ,
\end{eqnarray}
and  the above Lagrangean is obviously 
gauge invariant. We drop in the following 
the index $C$ since our results are the same for both $C$'s, and its 
inclusion should be obvious from the context. 
The mass and three-meson coupling 
terms can also be written with the
flavor indices $\alpha,\beta,\gamma $ explicit:

\begin{equation}
-\sum_\alpha  m_0^2\Phi^\alpha \Phi^{\alpha \dagger}+\sum_{\alpha\beta\gamma}
 (g_FC^{\alpha\beta\gamma}_- +g_DC^{\alpha\beta\gamma}_+)
  \Phi^{\alpha \dagger} \Phi^\beta \Phi^\gamma  \label{cterm}
\end{equation}
where $C^{\alpha\beta\gamma}_\pm$  is a set of Clebsch-Gordan-like 
numbers relating different coupling constants
\begin{equation}
C^{\alpha\beta\gamma}_\pm={\rm Tr}[{\bf \Lambda}^{\dagger \alpha }
{\bf \Lambda}^\beta{\bf \Lambda}^\gamma]_\pm \ .\label{trace}
\end{equation}
Chosing  the $\Lambda$ matrices to be the $\Lambda^{ij}$, which we defined
above
before  mixing ($\Omega =1$), these numbers are simply 0 or $\pm 1$ or 2
according to:
\begin{equation}
C^{ij,kl,mn}_\pm={\rm Tr}[{\bf \Lambda}^{ij\dagger}
{\bf \Lambda}^{kl}{\bf \Lambda}^{mn}]_\pm \label{trace1} 
=\delta_{jk}\delta_{lm}\delta_{ni} \pm \delta_{jm}\delta_{nk}\delta_{li}  
 \ .\label{traceij}                                                       
\end{equation}

Up til now flavor symmetry is assumed to be
exact and the bare nonet members have the same
mass $m_0$. But observe that the flavor fields 
 $\Phi^\alpha ={\rm Tr} [{\bf \Phi
\Lambda}^{\alpha \dagger } ]$, and the constants
$C^{\alpha\beta\gamma}_\pm $ of Eq.(\ref{trace}), are gauge invariant, 
and therefore we can 
break the flavor symmetry without destroying the 
gauge invariance, since each term in the sums 
in Eq. (\ref{cterm})  is gauge  invariant. 
Thus we can replace $m_0$, $g_F$ and $g_D$ by flavor dependent
(but gauge independent)  masses 
$m_\alpha $  and couplings $g_{\alpha\beta\gamma}$ in Eq.(\ref{cterm}) 
and still have a gauge invariant theory.
In fact, now the flavor breaking can be  spontaneous since gauge invariance
guarantees that any direction in the internal space chosen by
the flavor matrices is equivalent, as is obvious from Eq. (\ref{lam}).

Now considering
meson loops one must  renormalize the mass and coupling terms. 
The loops shift the masses and induce mixings between the flavorless
states  by terms $\Delta m_{\alpha\beta}^2 $  in the two-point functions
or the inverse popagators: 
\begin{equation}
P^{-1}_{\alpha\beta}(s)= m_0^2+\Delta m^2_{\alpha\beta}(s) -s  \ ,
\label{prop} \end{equation}
where
\begin{equation}
\Delta m_{\alpha\beta }^2(s) 
= \frac {g^2_F}{4\pi}\sum_{\gamma\delta}C^{\alpha\gamma\delta}_-
C^{\delta\gamma\beta}_-{\rm F}(s,m_\gamma ^2,m_\delta^2,\Lambda) 
+ \frac {g^2_D}{4\pi}\sum_{\gamma\delta}C^{\alpha\gamma\delta}_+
C^{\delta\gamma\beta}_+{\rm F}(s,m_\gamma ^2,m_\delta^2,\Lambda) \ .
\label{delta}  
\end{equation}

The zeroes of det$[P^{-1}_{\alpha\beta}(s)]$ 
determine the meson masses, which must by 
self-consistency be the same as the masses $m_\gamma$, 
$m_\delta$, which appear in the
threshold positions. In Eq.(\ref{prop})
 the constants $C^{\alpha\beta\gamma}_\pm $   contain the internal symmetry
dependence,
while the function F contains the kinematical dependence of the masses
in the loop. We return to models for the latter later below.
The constants $C^{\alpha\beta\gamma}_\pm$ satisfy the completeness relation
\begin{equation}
 \sum_{kl,mn} C^{ij,kl,mn}_\pm
C^{i'j',kl,mn}_\pm =N_f\delta_{ii'}\delta_{jj'}\pm \delta_{ij}\delta_{i'j'} \ , 
\end{equation}
as can easily be seen from their definition Eq.(\ref{trace}-\ref{trace1}).
In the case the solution is near the unmixed frame with $\Omega=1$, one can 
write Eq.(\ref{delta}) to a good approximation as
\begin{equation}
\Delta m_{ij,i'j'}^2(s) 
=\delta_{ii'}\delta_{jj'}\frac {g^2_F+g_D^2}{4\pi}\sum_{k}
{\rm F}(s,m_{ik}^2,m_{kj}^2,\Lambda)+ \delta_{ij}\delta_{i'j'}
\frac {g^2_D-g_F^2}{4\pi}{\rm F}(s,m_{ii'}^2,m_{jj'}^2,\Lambda) \ .
\label{deltaij} 
\end{equation}
where we have replaced $\alpha , \beta$, etc. 
by the quark-line indices $ij $ and $i'j' $ etc. 
In terms of quark-line diagrams
the first term represents a connected planar loop diagram,
and the second a disconnected diagram.
It is now easy to see that
if the symmetry is unbroken and all bare masses $=m_0$, then all
states with flavor get the same (normally negative) shift:
\begin{equation}
\Delta m^2 = \frac{g_D^2+g_F^2}{4\pi} N_f
{\rm F}(s,m_0^2+\Delta m^2,m_0^2+\Delta m^2 , \Lambda ) \label{deltamequal}
\end{equation}                                          
while for the flavorless states $u\bar u$, $d\bar d$,  $s\bar s$ 
one has an extra nondiagonal piece (for $N_f=3$):
\begin{equation}
 \Delta m^2_{ii,jj} =  \Bigl  [\frac{g_D^2+g_F^2}{4\pi}  
\left( \begin{array}{ccc} 3 & 0  & 0  \\
                          0 & 3  & 0  \\
                          0 & 0  & 3   \end{array}\right) + 
\frac{g_D^2-g_F^2}{4\pi}   \left( \begin{array}{ccc}  1 & 1 & 1 \\
                             1 & 1 & 1 \\
                             1 & 1 & 1     \end{array}\right) \Bigr ]
F(s,m^2_0+\Delta m^2,m^2_0+\Delta m^2 ,\Lambda ) \ .\label{deltagen}  
\end{equation}
 The second term
in (\ref{deltagen}) implies that the singlet is shifted differently from the
octet if $g_F \neq g_D$ (or if the $C=+$ mesons are not degenerate with those
with $C=-$). 
Thus when  summing over all  intermediate states $\gamma ,\delta $
 one self-consistently gets  the same mass shift 
function for all members of the octet when no symmetry breaking occurs.
 If $g_F=g_D$ the second nondiagonal term vanishes, and also 
the singlet is shifted equally, while if
 $g_F>g_D$\   the singlet will be  heavier than the octet. In general,
 there will
always exist one solution, which is exactly flavor symmetric, and which
for small couplings is  also the stable solution. But,  
for sufficiently large couplings the symmetric solution becomes unstable and
flavor symmetry is spontaneously broken!
 It is easiest to see how this  symmetry breaking mechanism works in 
the special case when there is no initial singlet-octet splitting and
$g_F=g_D=g$. Then only the first diagonal term proportional to $g_D^2+g_F^2$
contributes in Eq.(\ref{deltaij}).

 Let a small variation from the symmetric solution in
the threshold masses  be $\delta_{ij}=m_{ij}^2-m_0^2$. This
results in a shift $\delta^{\rm out}_{ij}$ in the pole
positions of Eq. (\ref{prop}). Self-consistency
of course requires $\delta^{\rm out}_{ij}=\delta_{ij}$ and stability
$\delta^{\rm out}_{ij}<\delta_{ij}$. 
The self-consistency condition 
requires that small deviations from the symmetric
solution must satisfy the equal spacing rule:
$\delta^{\rm out}_{ij}-\delta^{\rm out}_{ii}=(\delta^{\rm out}_{jj}-
\delta^{\rm out}_{ii})/2 $, while the instability condition $r=\delta^{\rm
out}_{ij}/\delta_{ij}>1$ can be written
\begin{equation}
N_f\frac {g_F^2+g_D^2}{4\pi}> [\frac {\partial\rm F}{\partial s}+\frac 
{\partial\rm  F}{\partial m_1^2}]^{-1}|_{s=m_1^2=m_2^2} \ . \label{instab}
\end{equation}
Eq.(\ref{instab}) is my most important result.  
Here the left hand side is positive for any reasonable function F,
which is essentially determined by the threshold behaviour.
In particular, assuming F to be given by its  unitarity cut,
Im(F)$\propto 
-k(s,m^2_1,m^2_2)/\sqrt s\ N(s)\theta (\Lambda-k)$, 
where $k$ is the 3-momentum in
the loop, and  $N(s)=1$ for simplicity, the left hand side of
Eq.(\ref{instab}) is given by the solid curve\footnote{
The two partial derivatives are here finite for  $\Lambda\to\infty$,
$\partial F/\partial s|_{s=m_1=m2}=1/\pi-2/\sqrt 27$, and
 $\partial F/\partial
m_1^2|_{s=m_1=m2}=1/\sqrt 27$. For large $\Lambda$ 
the bound in Eq.(\ref{instab}) thus approaches 7.94.} in Fig.2.
If one  has a P-wave behaviour at threshold,
Im(F)$=-k^3(s,m^2_1,m^2_2)/\sqrt s\
\theta (\Lambda-k)$ as for $\rho\to \pi\pi$, then the function F is much more 
sensitive to the threshold position, and consequently the bound is stricter by
more than an order
of magnitude  as seen by the dashed curve in Fig.2.
The coupling constants in Eq.(\ref{instab}) are  of course defined
before symmetry breaking, and are not the physical renormalized 
physical couplings after the breakdown. Still, it is interesting
to compare the bound with some typical 
physical coupling constants as done in Fig. 2.
As can be seen $g_{\rho\pi\pi} $ and $g_{\sigma\pi\pi}$ are well above the 
bound indicating that the spontaneous symmetry breaking under discussion
actually occurs in
Nature. The value used in Fig.2 for $g_{\sigma\pi\pi}$ (assuming
$m_\sigma\approx\Gamma_\sigma \approx 0.5$ GeV) is more like a lower bound
See ref.(\cite{NAT,NJL}). Now of course one can argue that for
such a large coupling the higher order loop diagrams  invalidate 
the approximation $N(s)=1$ (in the N/D terminology) made above. However,
the fact  that the bound is satisfied for all realistic cutoffs,
$\Lambda$ in an
effective theory, suggests strongly that it holds also in the real world.

Once the inequality (\ref{instab}) is satisfied one has an unstable situation.
Then, one must find out if there exists another solution
which is stable.                       
This  cannot be done analytically, since the
solution is highly nonperturbative and the function F is nonlinear
even in the simplest possible model. 
Once one is off the symmetric
situation, any mass splitting will feed into all other masses.
 In the first paper of 
Ref.(\cite{NAT3}) I did a numerical study of how the symmetry is 
broken, when $N_f=2$, $g_F=g_D=g$ is very large, and the 
 the function F determined by its unitarity cut assumed to be 
$ \propto -k(s,m_1^2,m_2^2) \sqrt s\  \theta (\Lambda -k)$, 
where $k $  is the  three-momentum
in the loop. It was shown that one finds for all $\Lambda$ 
a  stable flavor asymmetric 
solution which obeys approximately the equal spacing rule.

It is important to realize that once the equality is satisfied one has 
a discontinuous jump from the symmetric to the broken solution, - a small
increase in $g$ can give a large mass splitting within the multiplet.
For $N_f=3 $  it turns out that generally not the whole SU3 group
is broken, but that an SU2 subgroup remains unbroken, because of the
nonlinearities entering once the breaking appears. Thus, once
 the mass splittings 
$m_{s\bar s}-m_{s\bar d}\approx$ $m_{s\bar d} -m_{u\bar d}$ are nonzero,
then $u\bar d$, $d\bar u$, $u\bar u $  and $d\bar d $  remain degenerate.
This is like in the mechanical analogue above with collapsing elastic spheres 
to ellipsoides: O(3) symmetry is broken, 
but normally there  remains an exact O(2) symmetry around one 
of the principal axes.

 The situation is  more complicated when the singlet-octet splitting is
broken already for the symmetric solution or $g_F\neq g_D$.
Then for any  asymmetric solution 
also the mixing matrix $\Omega$ deviates from ideal mixing,   and
furthermore, $\Omega$ depends nonlinearily on all 
masses of the actual  solution. 
In which direction does the breaking then go?  
{\it A priori} the degenerate nonet mass  could  
split into  the 6 generally different masses of a nonet 
obeying $C$-symmetry, i.e. we have a 5-dimensional parameter space.
This looks  complicated,
but in the common case that the effect from singlet-octet splitting and
 $(g_F^2-g_D^2)/$$(g_F^2+g_D^2)$\ is  small compared to 
the spontaneous breaking between $s\bar s$, $s\bar u $ and $d\bar u $ 
masses, one can see analytically
the direction of the breaking. Considering the result above that an 
SU2 subgroup remains unbroken  while strange states are split from nonstrange
states one expects the solution to 
be near the ideally mixed frame, where isospin remains exact,
while  the $u\bar u$, $d\bar d $ and $s\bar s$ mix
through the off-diagonal terms in Eq.(\ref{deltaij}). Instead of
Eq.(\ref{deltagen}) one has a mass matrix of the form:
\begin{equation} \Delta m^2_{ii,jj} \propto
 \left( \begin{array}{ccc}  A &0&0\\0&A&0\\0&0&C  \end{array}\right) +
\frac {g_D^2-g_F^2}{g_F^2+g^2_D}  
 \left( \begin{array}{ccc}  a &a&b\\a&a&b\\b&b&c  
\end{array}\right) \to 
 \left( \begin{array}{ccc}  A &0&0\\0&A&0\\0&0&C  \end{array}\right) +
\frac {g_D^2-g_F^2}{g_F^2+g^2_D}  
 \left( \begin{array}{ccc}  0 &0&0\\0&2a&\sqrt 2 b\\0&\sqrt 2 b& c  
\end{array}\right)  \ ,
 \label{isos}
\end{equation}
where the second form is obtained after a rotation to pure  isospin
states. After further diagonalization one obtains mixing between
 $u\bar u+d\bar d $ and $s\bar s$.
  
In the second paper of Ref.(\cite{NAT3}) this mechanism was applied
to an SU6 model involving pseudoscalar and vector nonets, where the
nonet masses are degenerate before the spontaneous breaking.  
It was shown that  the symmetry
breaking was in the right direction and with a right  order of magnitude
in the  mass splittings.
Reasonable $\pi - K -\eta -\eta'$ and  $\rho -\omega - K^* -\phi$ 
splittings  was obtained when the scales were determined only by
the $\pi$ and $\rho$ masses.   

The main consequence of this spontaneous symmetry breaking is to
generate splittings between $s\bar s$, $s\bar d$ and $u\bar d$ states.
The same effect is obtained conventionally by inserting, by hand, into the
theory a heavier effective $s$-quark mass than  those of $u$, $d$.

Phenomenologically one can look at the instability I have discussed as
a phenomenon where, for sufficiently large coupling, 
a flavor symmetric meson cloud around a meson is unstable.
The meson clouds around  hadrons rearrange themselves through "virtual decay"
such that  stable, but flavor asymmetric  clouds are formed.

In the suggested symmetry breaking mechanism flavor 
states are {\it different} self-consistent {\it singlet} solutions within the
same internal (color) space. 
Flavor states do not carry color, nor do the Goldstone bosons carry flavor
as was assumed in some early attempts\cite{miya} to break flavor symmetry
spontaneously. The longitudinal and/or  scalar, confined 
gluons can thus be identified with the Goldstone bosons.
Color symmetry is then really the true unbroken symmetry behind flavor 
symmetry of strong interactions. When flavor symmetry
is a good approximation it only means 
that the solutions happen to be near each
others, because the bare states are (nearly) degenerate and the spontaneous 
symmetry breaking is absent or very weak.
In contrast to conventional spontaneous
symmetry breaking the new mechanism is clearly a quantum effect, where the 
symmetry breaking is in the solutions, not necessarily in the vacuum.
I believe a
better understanding of this spontaneous symmetry breaking
could throw light on the confinement problem, which I have not addressed.

I have discussed only three flavors and three colors. 
Of course we know there
are six flavors. In analogy with leptons 
three bare quarks should be massive because of weak interactions, 
but three may be (almost)
 massless. Extending my mechanism to 6 flavors 
all masses  will be renormalized and 
shifted from naive expectations. Thus, with this new mechanism,
the possibility remains open that with
three (nearly) massless and three massive bare quarks one might understand
the hadron mass spectrum.

\begin{figure}
\psfig{figure=fig1nat.epsf,width=8.5cm,height=2.4cm}                 
\vskip 0.2cm
\caption{ Classical spheres with fixed surface area collapsing to ellipsoids
when the internal pressure is reduced (See text).  The broken spherical shape
and the different the moments of inertia of the 
collapsing ellipsoids correspond to
the spontaneously broken global flavor symmetry, while the 
$x$-dependent O(3)-\-symmetry of rotating the ellipsoides correspond
to the unbroken  local  symmetry.}
\end{figure}                                                            

\begin{figure}
\psfig{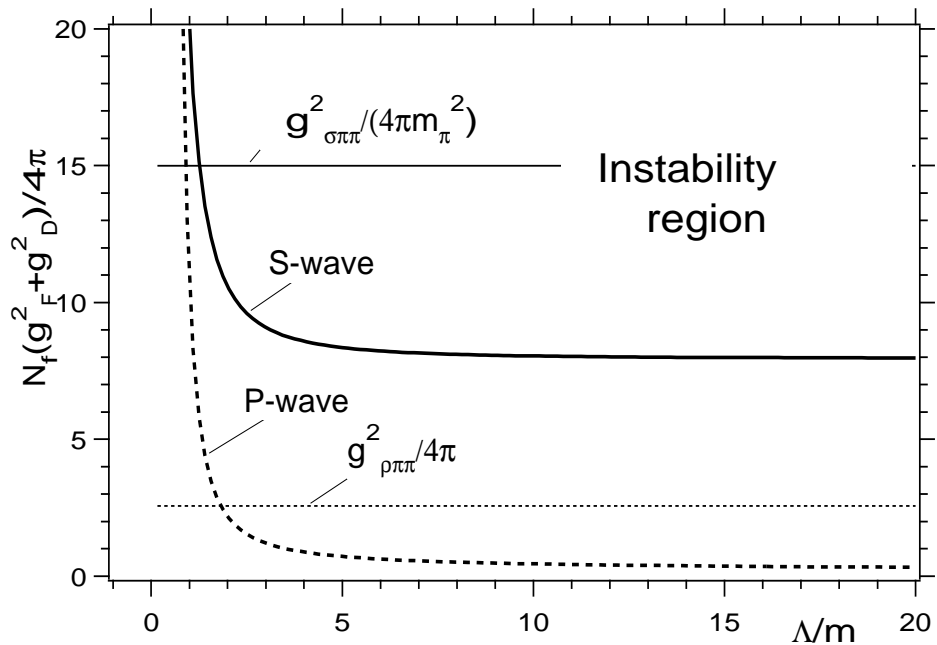}
\caption{ The instability limit on a dimensionless tri-meson
coupling constant. The solid curve is for an S-wave transition of a scalar
to two scalars, (like $\sigma\to \pi\pi$). 
The dashed curve is for a 
P-wave transition (like $\rho\to \pi\pi$).
 As shown the physical $g^2_{\rho  \pi \pi}/4\pi=2.5$ coupling constant, 
and a rough estimate for   
$g^2_{\sigma \pi \pi}/(4\pi m^2_\pi)\approx
15$ satisfy the instability condition for any reasonable value of the
cut off (e.g. in  $^3P_0$ decay models $\Lambda\approx 0.7$ GeV/c or
$\Lambda/ m_\pi \approx 5$).}
\end{figure}

\end{document}